\documentclass[prl,twocolumn,floatfix,showpacs]{revtex4-1}
\usepackage{graphicx}
\usepackage{amsmath}
\usepackage{amssymb}
\usepackage{color}
\usepackage{bm}
\usepackage{nccmath}

\begin{document}
\title{Universality and quantized response in bosonic nonfractionalized tunneling}
\author{Shaoyu Yin and Benjamin B\'eri}
\affiliation{School of Physics \& Astronomy, University of Birmingham, Edgbaston, Birmingham, B15 2TT, United Kingdom}
\date{July 2015}
\begin{abstract}
We show that tunneling involving bosonic wires and/or boson integer quantum Hall (bIQH) edges is characterized by universal features which are absent in their fermionic counterparts. Considering a pair of minimal geometries, we find 
a low energy enhancement and a universal high versus zero energy relation for the tunnel conductance that holds for all wire/bIQH edge combinations. Features distinguishing bIQH edges include a current imbalance to chemical potential bias ratio that is quantized despite the lack of conductance quantization in the bIQH edges themselves. The predicted phenomena require only initial states to be thermal and thus are well suited for tests with ultracold bosons forming wires and bIQH states. For the latter, we highlight a potential realization based on single component bosons in the recently observed Harper-Hofstadter bandstructure.

\end{abstract}

\maketitle

Tunneling setups are valuable probes of quantum matter with a scope that includes detecting strong correlation features, e.g., the suppression of tunneling in electronic quantum wires\cite{KFluttlett,*KFluttPRB}; topological aspects, e.g., zero bias features of Majorana fermions\cite{LawMaj,*Flencond,*wimmer2011quantum,bagrets2012class,*pikulin2012zero,*liu2012zero,*kells2012nz}; or even the combination of these, e.g., universal exponents and fractional charge for fractional quantum Hall (FQH) edge modes\cite{Wenedge1,*WenAW,*KaneFisherFraccharge,*ChangRMP}.

Most past work has focused on fermionic systems. Bosonic systems, however, are now attracting much interest, both due to theoretical predictions of novel,
nonfractionalized, symmetry protected topological (SPT) phases \cite{LuVishwanath,Chen2012,*Chen2013,*VishwanathSenthil,*SPTRevfn}, 
and experimental progress on ultracold atom systems such as the measurement of particle conductance\cite{Brantut12,*krinner2015observation} and the creation of topological bandstructures\cite{Aidel13,*Hirokazu13,*Aidel15,Jotzu14}.

\begin{figure}
\includegraphics[trim=0 0 -1cm -1cm,clip,width=\columnwidth]{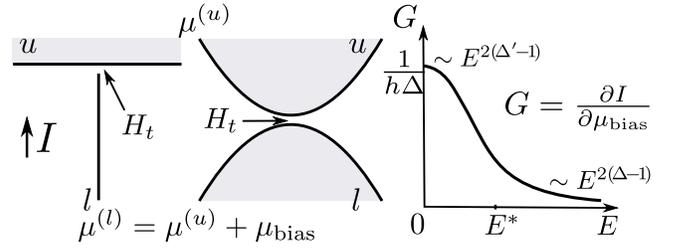}
\caption{Sketch of the tunneling setups and conductance. Left: T-junction. The lower ($l$) subsystem is a half-infinite wire, the upper ($u$) is a bIQH edge or a wire (or a spurious wire-like edge of a nontopological phase\cite{NTedgefn}). Shading indicates  the two-dimensional 
(2D) bulk.
Middle: QPC with bIQH edge or wire (wire-like) subsystems.  Our results also hold for the ``inverse QPC" with the 2D bulk between the  subsystems. Right:  
conductance versus low energy cutoff $E$ (e.g,  temperature, $\mu_\text{bias}$). $\Delta\!<\!1$, $\Delta^\prime\!>\!1$ are interaction dependent parameters. When $\Delta^\prime\! \neq \!\frac{1}{\Delta}$,  the Hall conductivity also enters $\Delta^\prime$. 
}
\label{fig:setup}
\end{figure}

In this Letter, we show that bosonic tunneling has 
striking universal features. Analyzing a variety of \mbox{setups} (Fig.~\ref{fig:setup}), we find that the tunneling conductance, in all cases,  is enhanced towards low energies, with the high and zero energy behavior being set by a single parameter. 
The T-junctions and quantum point contacts (QPCs) of Fig.~\ref{fig:setup} are the minimal setups admitting both  boson wire and boson integer quantum Hall (bIQH) edge\cite{SenthilLevinbIQH,LuVishwanath} subsystems: they provide both topological and nontopological paradigms for  bosonic nonfractionalized tunneling. The conductance features we find hold for all wire/bIQH edge 
combinations, 
and only require  wires/bIQH edges being clean and wires to have short range repulsive interactions, 
conditions naturally met with ultracold atoms.

This extent of the universality notwithstanding, we also show that the the topological nature of bIQH edge modes has clear tunnel transport signatures. 
The bIQH features we find include a current imbalance to chemical potential bias ratio  quantized  as   $\frac{\delta J_\rho}{\mu_\text{bias}}=\frac{q}{h}$ at low energies in a QPC setup  ($q\in\mathbb{Z}$ sets the bulk Hall conductivity, $\sigma=\frac{2q}{h}$\cite{Gquantfn}). This quantization is remarkable given that 
clean bIQH edges lack conductance quantization\cite{KaneFQHimlett,*KaneFQHim,MulliganFisher}.

It is instructive to compare our findings to their fermionic\cite{KFluttlett} and fractionalized\cite{Wenedge1} analogues and to a \mbox{$q=1$} bIQH result\cite{MulliganFisher}.
The enhancement we find is in stark 
contrast to the suppression for repulsive (and, for QPCs, also attractive) fermions and fermion/boson FQH edges in the same geometries. 
The contrast is also salient between hard-core bosons and free fermions: while  they are very 
similar in 1D, tunneling, 
just at a point, is already away enough from a pure 1D setting for bosonic tunneling enhancement to replace the energy independent free fermion behavior.  
A similar enhancement, characterizing  $q=1$ bIQH QPCs, was highlighted in Ref.~\onlinecite{MulliganFisher}. Our results show that this 
is part of a  universal behavior that arises for any $q$ and also for wires 
(or nontopological wire-like 1D modes).  Regarding $\frac{\delta J_\rho}{\mu_\text{bias}}$, it is interesting to note that it is \emph{not} quantized in fermionic integer quantum Hall QPCs: due to the chirality of the edge modes it is instead given by the (nonuniversal) tunnel conductance. 

We now turn to the analysis underlying our results. BIQH states are 2D SPT phases protected by particle number conservation; they support a pair of counterpropagating edge modes. Wires also respect this symmetry and have a pair of counterpropagating modes. This allows describing the two systems on the same footing using the K-matrix formulation developed for FQH systems\cite{WenZee}. The Hamiltonian of an edge or a wire is 
\begin{equation}
H_{\text 1D}=\frac{\hbar}{4\pi}\int dx \sum_{jk} \partial_x\phi_j V_{jk} \partial_x \phi_k
\label{eq:Hamiltonian}
\end{equation}
where $\phi_{1,2}$ are two bosonic fields satisfying\cite{SenthilLevinbIQH,LuVishwanath}
\begin{equation}
[\phi_{j}(x),\phi_{k}(y)]=i\pi K_{jk}^{-1}\text{sgn}(x-y),\quad K=\sigma_1 
\label{eq:Kmatrixcomm}
\end{equation}
with $\sigma_1$ the first Pauli matrix. In Eq.~\eqref{eq:Hamiltonian}, $V$ is a  symmetric positive definite matrix encoding the details of the confinement (or bandstructure in the wire case) and intermode interactions. The particle density (relative to a background value)
is given by $n_{\rho}(x)=\frac{1}{2\pi}\mathbf{q}^{T}\partial_{x}\boldsymbol{\phi}(x)$, with the integer vector $\mathbf{q}^{T}=(1,q)$. The key  
difference between wires and bIQH edges is in $\mathbf{q}$: $q=0$ for the former while $q\neq0$ for the latter\cite{LuVishwanath}. 
This distinction also leads to different 
operators creating/removing a given number of particles (or ``charge"). We have $\psi_{\boldsymbol{\lambda}}^\dagger(x)\sim e^{-i\boldsymbol{\lambda}^{T}\boldsymbol{\phi}}$ with an integer vector $\boldsymbol{\lambda}$, creating charge $Q_{\boldsymbol{\lambda}}=\mathbf{q}^{T}K^{-1}\boldsymbol{\lambda}$. Note that $Q_{\boldsymbol{\lambda}}$ is integer; the theory has no fractional quasiparticles. In the bIQH cases, without loss of generality we assume $q>0$, and take $V_{11}<V_{22}$. While the latter is an  immaterial choice 
for $q=1$\cite{Vmatfn}, for $q>1$ it follows from requiring the dominant edge mode interaction to be through the particle density.

For wires, the reflection symmetric\cite{Giabook}  \mbox{$V_{12}\!=\!0$}  case gives the usual Luttinger liquid with density \mbox{$n_\rho(x)\!=\!\frac{\partial_x \phi_1}{2\pi}$}, phase $\phi_2$,
and Luttinger parameter \mbox{$g\!=\!\frac{1}{2}\sqrt{\frac{V_{22}}{V_{11}}}$}. 
For the short range repulsive interactions we consider we have $g\!\geq\! 1$, with $g\!=\!1$ being the hard-core limit\cite{Giabook}.
In T-junctions, for the half-infinite wires ending at $x\!=\!0$ we have $\phi_1(x\!\rightarrow\!0)=\!0$\cite{Nayak99,*Oshikawa06}. 
For the  wire-like nontopological cases 
in Fig.~\ref{fig:setup}, $V_{12}\neq 0$ will be allowed.

The tunneling between the upper ($u$) and lower ($l$) subsystems is implemented by the coupling 
\begin{equation}
H_t=\sum_{\boldsymbol{\lambda}_u,\boldsymbol{\lambda}_l} t_{\boldsymbol{\lambda}_u,\boldsymbol{\lambda}_l}\ e^{i(\boldsymbol{\lambda}_l^{T}\boldsymbol{\phi}^{(l)}-\boldsymbol{\lambda}_u^{T}\boldsymbol{\phi}^{(u)})}+\text{h.c.}
\label{eq:Htun}
\end{equation}
with the charge conservation constraint $Q_{\boldsymbol{\lambda}_u}+Q_{\boldsymbol{\lambda}_l}=0$ and the fields taken at $x=0$. Treating $H_t$ as a  
perturbation, the relevance of the different terms upon lowering the energy scale is determined by their scaling dimension $\Delta_{\boldsymbol{\lambda}_u,\boldsymbol{\lambda}_l}$: under a renormalization group (RG) elimination of the window $(\frac{D_0}{b}, D_0)$ below the bare high energy cutoff $D_0$, the couplings flow as $t_{\boldsymbol{\lambda}_u,\boldsymbol{\lambda}_l}=t^{(0)}_{\boldsymbol{\lambda}_u,\boldsymbol{\lambda}_l}b^{1-\Delta_{\boldsymbol{\lambda}_u,\boldsymbol{\lambda}_l}}$. ($D_0$ is set, e.g, by the 
background density
for wires, or the gap for bIQH states.) We have $\Delta_{\boldsymbol{\lambda}_u,\boldsymbol{\lambda}_l}=\Delta_{\boldsymbol{\lambda}_{u}}+\Delta_{\boldsymbol{\lambda}_{l}}$ where $\Delta_{\boldsymbol{\lambda}}$ is the scaling dimension of an $e^{\pm i\boldsymbol{\lambda}^{T}\boldsymbol{\phi}}$ factor in Eq.~\eqref{eq:Htun}. 
For the end of a half infinite wire this is $\Delta_{\boldsymbol{\lambda}}=\frac{Q_{\boldsymbol{\lambda}}^2}{2g}$, while for a bIQH edge or the bulk of a wire 
$\Delta_{\boldsymbol{\lambda}}=\frac{1}{2}\sqrt{\frac{V_{22}}{V_{11}}}\lambda_{1}^{2}+\frac{1}{2}\sqrt{\frac{V_{11}}{V_{22}}}(Q_{\boldsymbol{\lambda}}-q\lambda_{1})^{2}$. 
The dominant term (with the smallest $\Delta_{\boldsymbol{\lambda}_u,\boldsymbol{\lambda}_l}$)
is the single particle process $H_{t_{0}}=t_{0} \exp[i(\phi_{2}^{(l)}-\phi_{2}^{(u)})]+\text{h.c.}$. The scaling dimension $\Delta=\Delta_u+\Delta_l$ is  
$\Delta=\frac{1}{2}\sqrt{\frac{V^{(u)}_{11}}{V^{(u)}_{22}}}+\frac{1}{2g_l}$ for T-junctions, and $\Delta=\frac{1}{2}\sqrt{\frac{V^{(u)}_{11}}{V^{(u)}_{22}}}+\frac{1}{2}\sqrt{\frac{V^{(l)}_{11}}{V^{(l)}_{22}}}$ for QPCs. Importantly, $\Delta<1$: 
the coupling increases under RG as the energy scale is lowered. This translates to the high energy tail of the conductance enhancement announced above (see also Fig.~\ref{fig:setup}),   $G\sim E^{2(\Delta-1)}$ 
for  $E\gg E^*= D_0\left(\frac{|t_{0}^{(0)}|}{D_0}\right)^{\frac{1}{1-\Delta}}$, where $E$ is the infrared  
cutoff, e.g,  temperature $T$ or 
bias $\mu_\text{bias}$. (The power laws here and below hold for $E$ much larger than other infrared energy scales.)

That the high energy conductance has the same schematic form for all junctions is just a consequence of Fermi's golden rule. It is less expected that, in all cases, $\Delta$ also sets 
the zero energy limit, $G=\frac{1}{h\Delta}$. 
We now establish this result, assuming that the zero energy physics is governed by $H_{t_0}$ at strong coupling. 

\begin{figure}
\includegraphics[trim=0 0 0 -1cm,clip,width=\columnwidth]{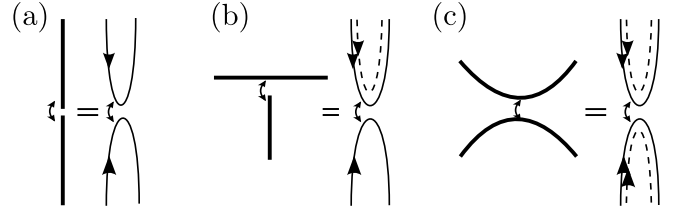}
\caption{Mapping T-junctions and QPCs to linear junctions. Diagram (a) shows the unfolding for linear junctions; diagrams (b) and (c) show the 
transformations for T-junctions and QPCs. Dashed lines indicate the 
decoupled modes. 
}
\label{fig:mapping}
\end{figure}

The key observation is that with $H_{t_0}$ only, we can relate our setups to a junction between two half-infinite wires (see Fig.~\ref{fig:mapping}). In that case, $\Delta=\frac{1}{2g_u}+\frac{1}{2g_l}$ and  $G=\frac{1}{h\Delta}$ follows from results\cite{KFluttlett,ChamonFradkin,*ChangYu12} on fermion Luttinger liquids with $\Delta<1$, upon observing that the wire Hamiltonian, the tunnel coupling and the correlation functions (current-current correlators calculated, e.g., in the lower subsystem) underlying $G$ at zero energy are defined by the same expressions in the boson and (bosonized) fermion cases. Another useful viewpoint  
is one where  the left and right moving modes of half-infinite wires are unfolded\cite{Nayak99} so that the junction consists of two right movers $\phi^{(u,l)}_R$ on the full line, coupled at $x=0$. 
In terms of these, the wire Hamiltonians are
$H_{\text 1D}^{(u,l)}=\frac{\hbar v^{(u,l)}}{4\pi}\int dx (\partial_x \phi^{(u,l)}_R)^2$, with $[\phi_R^{(j)}(x),\phi_R^{(k)}(y)]=i\pi \delta_{jk}\text{sgn}(x-y)$, and the coupling is $H_{t_0}= t_0\exp[i(\sqrt{2\Delta_u}\phi_R^{(u)}-\sqrt{2\Delta_l}\phi_R^{(l)})]+\text{h.c}$. 

Turning now to T-junctions, one can flip the left mover component on the $u$ (bIQH or wire) side and rescale $x$  to formally treat $\phi_2^{(u)}$ as the combination of two right movers with the same velocities. One can then introduce this combination as a new right-mover field $\phi^{(u)}_{R1}$ (and the orthogonal combination as $\phi^{(u)}_{R2}$). In terms of these, the $u$ side Hamiltonian is $H_{\text 1D}^{(u)}=\frac{\hbar v}{4\pi}\sum_{\alpha=1,2}\int dx (\partial_x \phi_{R\alpha}^{(u)})^2$, with $[\phi^{(u)}_{R\alpha}(x),\phi^{(u)}_{R\beta}(y)]=i\pi \delta_{\alpha\beta}\text{sgn}(x-y)$, and the coupling is  $H_{t_0}= t_0\exp[i(\sqrt{2\Delta_u}\phi_{R1}^{(u)}-\sqrt{2\Delta_l}\phi_R^{(l)})]+\text{h.c}$ where $\phi_R^{(l)}$ is the unfolded right mover on the $l$ (half-infinite wire) side. 
The field $\phi^{(u)}_{R2}$  decouples.  
From the $l$ side, 
the $u$ side is seen as another half-infinite wire with Luttinger parameter set by $\Delta_u$. Calculating the zero energy conductance from $l$ side current-current correlators, we therefore find $G=\frac{1}{h\Delta}$ (with $\Delta$ for the T-junction). Of course, the conductance should be the same if it is calculated from $u$ side correlators. This immediately shows that $G=\frac{1}{h\Delta}$ also holds for QPCs: performing now the same steps on the $l$ side, from the perspective of the $u$ side a QPC looks the same as a T-junction with the $l$ side  Luttinger parameter set by $\Delta_l$. This completes our proof that $G=\frac{1}{h\Delta}$ holds for all junctions at zero energy.

The analysis of the low energy behavior is completed by considering the scaling dimensions $\Delta^\prime$ of perturbations near strong coupling. These are obtained taking the $x\rightarrow 0$ limit of
charge conserving operators, subject to the boundary condition implied by $t_0\rightarrow \infty$\cite{Nayak99,BeriTK,*MultiCoulMaj,*MajKlein,*ABET1}. Starting with linear junctions, for completeness, the leading perturbation has $\Delta^\prime= \frac{1}{\Delta}$. 
For T-junctions we also find $\Delta^\prime=\frac{1}{\Delta}$ unless $\gamma=\sqrt{\frac{V^{(u)}_{22}}{V^{(u)}_{11}}}$ satisfies $g_l-\sqrt{g_l^2-q^2}<\gamma<g_l+\sqrt{g_l^2-q^2}$ (for $g_l>q$), in which case $\Delta^\prime=\frac{q^2+\gamma^2+2\gamma g_l}{2(\gamma+g_l)}$. For QPCs, taking the same interactions on the $u/l$ sides for brevity, we find $\Delta^\prime= \frac{1}{\Delta}$ for $\frac{1}{q}<\sqrt{\frac{V_{11}}{V_{22}}}<1$ and $\Delta^\prime=\frac{3+\Delta^2q^2}{4\Delta}$ for $\sqrt{\frac{V_{11}}{V_{22}}}<\frac{1}{q}$ (with $\frac{1}{q}=\infty$ for $q=0$). Importantly, for the interactions we consider  $\Delta^\prime>1$: the strong coupling fixed point is stable. (This also holds for QPCs with unequal $u/l$ side interactions.) 
The dimension $\Delta^\prime$ 
governs the
corrections to the zero energy conductance, $\delta G \sim E^{2(\Delta^\prime-1)}$ ($E\ll E^*$). 
It is interesting to note that for $g_u=g_l$ boson wire T-junctions and QPCs the low energy $G$ also agrees with that of multi-junctions of semi-infinite wires\cite{AkiyukiOshikawaDemler,*crampe2013quantum} and fermionic topological Kondo systems\cite{BeriTK} related to them by bosonization.

The possibility of a power law $\delta G \sim E^{2(\Delta^\prime-1)}$ with $\Delta^\prime\neq \frac{1}{\Delta}$ is the first feature that may be used to detect bIQH edge modes via tunneling: if such a $\Delta^\prime$ is
measured, $q$ can be extracted from it
using $\Delta$ for a symmetric QPC, or $\Delta$ and $g_l$ for a T-junction, with $\Delta$ given by $G$ at zero or high energy, and $g_l$ obtained\cite{Giabook} 
either via data on the microscopic $l$ side interactions, or measurements, e.g., of $l$ side density correlations. 
Note that in T-junctions, $l$ side interactions can always be tuned to reach $\Delta^\prime\neq \frac{1}{\Delta}$.

While this is already a well defined bIQH feature, as we explain below, tunneling setups also provide signatures that are more qualitative.
By solid-state analogy, the Hall resistance $R_H$ in a four terminal arrangement would be a natural candidate. 
With cold atoms, however, measuring the Hall voltage $V_H$
and having thermalized edge modes for its clear interpretation may be challenging. 
Surprisingly, even if these issues are resolved, 
$R_H$ turns out to vanish for a clean edge at zero temperature 
due to $H_{t_0}$ remaining the only coupling: firstly, due to Eq.~\eqref{eq:Kmatrixcomm}, $H_{t_0}$ tunnels only into mode $\phi_1$, thus measuring $V_H$ probes $\mu_1$ (where $\mu_j$ is the chemical potential of mode $j$) and the $\phi_2$ current  satisfies $\partial_x J_2=0$ in a steady state. Secondly, in a thermal state we have\cite{Kanecontacts} 
$J_2=\mu_1/h$.
Therefore, $\mu_1$ is constant along the whole edge, implying that $V_H$ and thus $R_H$ vanishes.

\begin{figure}
\includegraphics[trim=0 0 0 0cm,clip,width=\columnwidth]{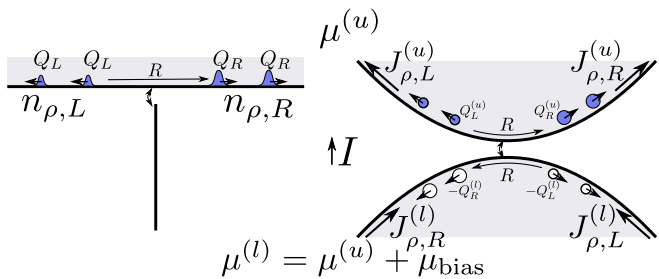}
\caption{Charge densities $n_{\rho,R/L}$ and currents $J_{\rho,R/L}$ underlying predictions 
Eqs.~\eqref{eq:chargeimbalance} and \eqref{eq:Jquant}. We measure $n_{\rho,R/L}$ and $J_{\rho,R/L}$ relative to their equilibrium values at $\mu^{(u/l)}$. The right moving direction ($R$ arrows) is taken relative to the Hall state chirality.  For $\delta J_{\rho}\!=\!J^{(u)}_{\rho,R}\!-\!J^{(l)}_{\rho,L}$ we find $\frac{{\delta J_\rho}}{\mu_\text{bias}}=\frac{q}{h}$ at low energies.  
The intuitive picture is in terms of charge $Q^{(u/l)}_{R/L}$ solitons. 
}
\label{fig:rhoJ}
\end{figure}

In fact, the observation that $H_{t_0}$ tunnels only to $\phi_1$ explains all the bIQH-boson wire similarities we encountered: 
if tunnel junctions are the only sources of data (e.g., via $G$ or as voltage probes) one effectively probes only
the $\phi_1$ component of $\mathbf{q}$ which is $q$ independent. 
A bIQH state being a Hall system, a transport current 
should however, regardless of details, induce some charge density $\delta n_\rho$ and current $\delta J_\rho$ imbalance between opposite edges which should be $q$ dependent.  
Not requiring voltage detection, nor thermalized edges, these observables are natural alternatives to $R_H$ for cold atom settings. 
As we now show, tunneling setups are particularly advantageous for 
measuring $\delta n_\rho$ and $\delta J_\rho$
as  nonuniversalities due to edge mode interactions can be 
eliminated using $\Delta$ (and $g_l$ in T-junctions). 
The key quantities are summarised in Fig.~\ref{fig:rhoJ}.  
The equations of motion relate the currents $\mathbf{J}$ and densities $\mathbf{n}=\frac{\partial_x \boldsymbol{\phi}}{2\pi}$ of the two edge modes as $\mathbf{J}=K^{-1}V \mathbf{n}$\cite{Kanecontacts}.
When $H_{t_0}$ dominates, in a steady state we have $\partial_x J_1=I\delta(x)$ and $\partial_x J_2=0$ up to small corrections. Integrating across the contact  we find
\begin{equation}
\frac{\delta n_\rho}{I}=\frac{1+q\sqrt{\frac{V_{11}}{V_{22}}}}{2v_{R}}-\frac{1-q\sqrt{\frac{V_{11}}{V_{22}}}}{2v_{L}}
\label{eq:chargeimbalance}
\end{equation}
for the charge imbalance $\delta n_\rho=n_{\rho,R}-n_{\rho,L}$ between right/left edge segments. Here $v_{R/L}=\sqrt{V_{11}V_{22}}\pm V_{12}$ are the velocities of right/left movers, measuring which should be feasible in cold atom experiments\cite{Neutfn}. The deviation of $\delta n_\rho$ from $\frac{1}{2}(v_R^{-1}-v_L^{-1})$ is a qualitative bIQH signature.
In a T-junction we have $\sqrt{\frac{V_{11}}{V_{22}}}=2\Delta-g_l^{-1}$ thus $q$ can be extracted from  $\frac{\delta n_\rho}{I}$ using $v_{R/L}$, $\Delta$ and $g_l$. 

Eq.~\eqref{eq:chargeimbalance} can be also derived using an intuitive picture noting that $Q_{R/L}=\frac{1\pm q\sqrt{\frac{V_{11}}{V_{22}}}}{2}$ is 
the charge of right/left moving solitons that arise due to injecting unit charge into $\phi_1$ (i.e., a single $H_{t_0}$ transport event). Identifying $I$ as the injection rate, $n_{\rho,R/L}=\frac{Q_{R/L}}{v_{R/L}}I$ is the  
charge density to the right/left of the contact, thus explaining Eq.~\eqref{eq:chargeimbalance}. Measuring $\frac{n_{\rho,R/L}^{(u/l)}}{I}$, $v_{R/L}$ and $\Delta$, one can also extract $q$ using a QPC. QPCs however also provide a more elegant option via the current $J_{\rho,L/R}=v_{L/R} n_{\rho,L/R}$. 
In terms of $\delta J_\rho=J^{(u)}_{\rho,R}-J^{(l)}_{\rho,L}$
we find $\frac{{\delta J_\rho}}{I}=q\Delta$ or simply
\begin{equation}
\frac{{\delta J_\rho}}{\mu_\text{bias}}=\frac{q}{h}, 
\label{eq:Jquant}
\end{equation}
the quantization announced above, which holds at low energies, up to small power law corrections. Note that Eq.~\eqref{eq:Jquant} does not require a symmetric QPC.

We close with some considerations on testing our predictions in ultracold atomic systems. Our results require a steady state but only need initial states to be thermal.  One can therefore envisage testing them in a scheme inspired by
Ref.~\onlinecite{Brantut12}: 
with $H_t$ off, preparing the $u$ and $l$ subsystems with a small 
$\mu_\text{bias}$ between them and then turning $H_t$ on to let the full system evolve. 
In the steady state, we expect\cite{Einhellinger12,*Sabetta13} 
$I=G[\max(T,\frac{\hbar}{t_s})]\mu_\text{bias}$ for $\mu_\text{bias}, \min(T,\frac{\hbar}{t_s})\ll \max(T,\frac{\hbar}{t_s})$, where $t_s$ is the steady state duration. The imbalance $\delta n_\rho$ can be measured by standard methods for detecting particle density. The recent demonstration of detecting local lattice currents\cite{atala2014observation} also provides prospects for observing $\delta J_\rho$.

The $\delta n_\rho$ and $\delta J_\rho$ predictions require a bIQH realization. 
This was originally proposed\cite{SenthilLevinbIQH} and 
later numerically demonstrated\cite{Shunsuke13,*Regnault13,*HaiJain13,*Grass14}  
for two-component bosons in the lowest Landau level. Lattice systems of current experiments\cite{Aidel13,Jotzu14}, however, may provide a bIQH option already with single-component particles. This is easiest seen for $q\!=\!1$: 
$K\!=\!\sigma_1$ and $\mathbf{q}\!=\!(1,1)$ is in the composite fermion 
series\cite{JainCF} for bosons in a periodic potential\cite{KolRead,MollerCooper}, corresponding to the case when composite fermions fill bands of total Chern number $|C|\!=\!2$ (as $K$ is $2\!\times\! 2$) and bind a single flux quantum [transforming $K_0\!=\!-\openone_2$ of filled bands of Chern number -2  to $K\!=\!-\openone_2\! +\! \bigl(\begin{smallmatrix}
1 & 1\\
1 & 1
\end{smallmatrix}\bigr)$]\cite{WenZee}. Interestingly, this state was
numerically observed\cite{MollerCooper} in an FQH study of the Harper-Hofstadter model, providing the earliest instance of a bIQH state (see also Ref.~\onlinecite{MollerCooper15}). 
FQH results on the same model also suggest that well defined edge modes may be possible under realistic conditions\cite{Kjall12}. 
The cold atom realisation of the Harper-Hofstadter bandstructure\cite{Aidel13} is thus particularly promising from the bIQH perspective. 
Recently, bIQH states were also predicted in optical flux lattices\cite{Sterdyniak15} and in 
models with correlated hopping\cite{Pollmann15}.

In conclusion, we have shown that bosonic nonfractionalized tunneling 
displays emblematic universal features. Tunneling is enhanced upon lowering the energy, with the high energy power law $G\sim E^{2(\Delta-1)}$  tied to the zero energy limit $G=\frac{1}{h\Delta}$. This universal relation holds for all the interactions and geometries we considered, and is valid regardless of the topological nature of the modes. While wires and bIQH edges 
are thus remarkably similar as far as the tunneling conductance is concerned, we have also shown how bIQH states are revealed in tunneling transport via low energy power laws, the charge imbalance 
Eq.~\eqref{eq:chargeimbalance}, and the quantization $\frac{{\delta J_\rho}}{\mu_\text{bias}}=\frac{q}{h}$. Our predictions, requiring only initial states to be thermal, are well suited for cold atom experiments, where, as we noted,  
there is progress towards concrete bIQH host bandstructures. 
It will be interesting to explore generalizations of our results to other SPT phases.

We acknowledge useful discussions with N. R. Cooper and J. M. F. Gunn. This work was supported by a Royal Society and a Birmingham  Research Fellowship.



\begin{thebibliography}{59}
\bibitem[{Gqu()}]{Gquantfn}
\bibinfo{note}{We define conductivity and conductance as particle current
  against chemical potential bias, in terms of which the conductance quantum is
  $1/h$.}\relax
\bibitem[{Vma()}]{Vmatfn}
\bibinfo{note}{In the $q=1$ case our choice only implies working away from the
  special $V_{11}=V_{22}$ point, reaching which would require careful fine
  tuning of confinement details.}\relax
\bibitem[{Neu()}]{Neutfn}
\bibinfo{note}{If $\frac{V_{11}}{V_{22}}=q^{-2}$, the left moving mode, akin to
  an edge dipole mode, carries no particle density. While this makes detecting
  $v_L$ challenging, the density propagating only to the right provides a
  qualitative bIQH signature.}\relax
\end{thebibliography}
\end{document}